\RequirePackage{lineno} 
\documentclass[aps,prc,superscriptaddress,showpacs,floatfix,nofootinbib,onecolumn]{revtex4}
\usepackage{graphicx} 
\usepackage{float} 
\usepackage{xfrac} 
\usepackage{amssymb}
\usepackage{url}
\usepackage{harpoon}
\usepackage{amsmath}
\usepackage{MnSymbol}
\usepackage{multirow}
\usepackage[colorlinks,breaklinks]{hyperref}
\hypersetup{linkcolor=blue,citecolor=blue,filecolor=black,urlcolor=blue}

\newcommand{\sub}[1]{_{\mathrm{#1}}}
\newcommand{\su}[1]{^{\mathrm{#1}}}
\newcommand{\lr}[1]{\left\langle #1\right\rangle}
\newcommand{\lrp}[1]{\left( #1\right)}
\newcommand{\pT} {\ensuremath{p_{\mathrm{T}}}}

\newcommand{\Dphi}{\mbox{$\Delta \phi$}}
\newcommand{\Deta}{\mbox{$\Delta \eta$}}

\newcommand{\npartf}{N_{\mathrm {part}}^{\mathrm{F}}}
\newcommand{\npartb}{N_{\mathrm {part}}^{\mathrm{B}}}
\newcommand{\npart}{N_{\mathrm {part}}}
\newcommand{\nspecf}{N_{\mathrm {spec}}^{\mathrm{F}}}
\newcommand{\nspecb}{N_{\mathrm {spec}}^{\mathrm{B}}}

\newcommand{\nneuf}{N_{\mathrm {neu}}^{\mathrm{F}}}
\newcommand{\nneub}{N_{\mathrm {neu}}^{\mathrm{B}}}
\newcommand{\nneu}{N_{\mathrm {neu}}}

\begin{document}
\title{Forward-backward multiplicity fluctuation and longitudinal harmonics in high-energy nuclear collisions}
\newcommand{\sunysb}{Department of Chemistry, Stony Brook University, Stony Brook, NY 11794, USA}
\newcommand{\bnl}{Physics Department, Brookhaven National Laboratory, Upton, NY 11796, USA}
\author{Jiangyong Jia}\email[Correspond to\ ]{jjia@bnl.gov}\affiliation{\sunysb}\affiliation{\bnl}
\author{Sooraj Radhakrishnan}\affiliation{\sunysb}
\author{Mingliang Zhou}\email[Correspond to\ ]{mingliang.zhou@stonybrook.edu}\affiliation{\sunysb}
\begin{abstract}
An analysis method is proposed to study the forward-backward (FB) multiplicity fluctuation in high-energy nuclear collisions, built on the earlier work of Bzdak and Teaney. The method allows the decomposition of the centrality dependence of average multiplicity from the dynamical event-by-event (EbyE) fluctuation of multiplicity in pseudorapidity. Application of the method to AMPT and HIJING models shows that the long-range component of the FB correlation is captured by a few longitudinal harmonics, with the first component driven by the asymmetry in the number of participating nucleons in the two colliding nuclei. The higher-order longitudinal harmonics are found to be strongly damped in AMPT compare to HIJING, due to weaker short-range correlations as well as the final-state effects present in the AMPT model. Two-particle pseudorapidity correlation reveals interesting charge-dependent short-range structures that are absent in HIJING model. The proposed method opens an avenue to elucidate the particle production mechanism and early time dynamics in heavy-ion collisions. Future analysis directions and prospects of using the pseudorapidity correlation function to understand the centrality bias in $p$+$p$, $p$+A and A+A collisions are discussed. 
\end{abstract}
 \maketitle 

\section{Introduction} 
\label{sec:1}
Heavy-ion collisions at RHIC and LHC have two defining characteristics which are the focus of many studies: 1) large density fluctuations in the initial state of the collisions that varies event to event, and 2) the rapid formation of a strongly coupled quark gluon plasma that expands hydrodynamically with very low specific viscosity. The latter characteristic leads to a very efficient transfer of the initial density fluctuations into the final-state collective flow correlations in momentum space. Conversely, experimental measurements of the these correlations provide a window into the the space-time picture of the collective expansion as well as the medium properties that drives the expansion. The measurement of harmonic flow coefficients $v_n$~\cite{Adare:2011tg,ALICE:2011ab,Aad:2012bu,Chatrchyan:2013kba} and their event-by-event (EbyE) fluctuations~\cite{Aad:2013xma,Aad:2014fla,Aad:2015lwa} has placed important constraints on the shear viscosity and density fluctuations in the initial state~\cite{Luzum:2013yya,Gale:2013da,Heinz:2013th,Jia:2014jca}.

Recently, similar ideas have been proposed to study the initial state density fluctuations in the longitudinal direction~\cite{Bozek:2010vz,Bzdak:2012tp,Jia:2014ysa,Bhalerao:2014mua}. These longitudinal fluctuations directly seed the entropy production at very early time of the collisions, well before the onset of the collective flow, and appear as correlations of the multiplicity of produced particles separated in rapidity. For example, EbyE difference between the number of nucleon participants in the target and the projectile, $\npartf$ and $\npartb$ may result in a long-range asymmetry of the fireball~\cite{Bialas:2011bz,Bzdak:2012tp,Jia:2014ysa}; the fluctuation of emission profile among participants may lead to higher-order shape fluctuations in rapidity~\cite{Bzdak:2012tp,Jia:2014vja} (assuming that the emission sources for particle production can be associated with individual wounded nucleons). On the other hand, short-range correlations can also be generated dynamically including resonance decay, jet fragmentation and Bose-Einstein correlations. These correlations are typically localized over a smaller range of the $\eta$ and can be sensitive to final-state effects. The longitudinal multiplicity fluctuations, when coupled with the collective transverse expansion, also lead to rapidity-dependent EbyE fluctuations of magnitude and the phase of harmonic flow~\cite{Bozek:2010vz,Jia:2014ysa,Pang:2014pxa,Khachatryan:2015oea}.

Most previous studies of the longitudinal multiplicity correlation are limited to two rapidity windows symmetric around the center-of-mass of the collision system, commonly known as forward-backward (FB) correlations~\cite{Bialas:2010zb,Vechernin:2013vpa}. They have been measured experimentally in $e^+e^-$~\cite{Braunschweig:1989bp}, $p+p$~\cite{Ansorge:1988fg,Uhlig:1977dc,ATLAS:2012as,Adam:2015mya}, $p+\bar{p}$~\cite{Alexopoulos:1995ft} and A+A~\cite{Back:2006id,Abelev:2009ag} collisions where significant FB asymmetric component has been identified. Recently, Refs.~\cite{Bzdak:2012tp,Bhalerao:2014mua} generalized the study of the shape of the rapidity fluctuation by decompose it into Chebyshev polynomials or into principle components, with each mode representing the different components of the measured FB correlation. In this paper, we propose a single-particle method that obtains these shape components directly from each event, as well as a two-particle correlation method that gives the ensemble RMS-average of these shape components. We apply the method to HIJING~\cite{Gyulassy:1994ew} and AMPT~\cite{Lin:2004en} models and successfully extract the different shape components of the multiplicity fluctuation. The first component is found to be directly related to the long-range asymmetry of the fireball, while the higher-order components are more related to the short-range correlations. The extracted components are also found to be dampened by the final-state interactions. Therefore our method can be used for systematic study of the longitudinal dynamics in heavy-ion collisions.

The structure of the paper is as follows. The next section introduces the method and relates to previous observables. Sections~\ref{sec:sp} and \ref{sec:spec} show the properties of the longitudinal shape components extracted from HIJING and AMPT models. The meaning of the first few components are discussed within the context of a simple wounded-nucleon and particle emission model, and their relations to initial density fluctuations are clarified. Section~\ref{sec:2pc} compares between the single-particle and correlation methods, and a procedure is introduced to further decouple residual centrality dependence from the dynamical FB correlations in the correlation function. Section~\ref{sec:dis} discusses new analyses enabled by the method, as well as its potential application for understanding the centrality bias effects. 
\section{The method} 
\label{sec:method}
The FB correlation can be quantified by two-particle correlation (2PC) function, see for example Ref.~\cite{Vechernin:2013vpa}:
\begin{eqnarray}
\label{eq:1}
C(\eta_1,\eta_2) &=& \frac{\left\langle N(\eta_1) N(\eta_2)\right\rangle -\left\langle N(\eta_1)\right\rangle\delta(\eta_1-\eta_2)}{\left\langle N(\eta_1)\right\rangle\left\langle N(\eta_2)\right\rangle}
\end{eqnarray}
where the $ N(\eta)\equiv dN/d\eta$ is multiplicity density distribution in pseudorapidity in one event, the average is over the event ensemble, e.g. events within a given centrality class. In experimental analysis, correlation function is usually normalized to have an average value of one. The second term in the numerator explicitly removes the self-correlation contribution, i.e. one should not correlate a particle with itself. This term is usually dropped in the standard notation, since condition $\eta_1\neq \eta_2$ is implicitly assumed, but it is important in our discussion for reasons that will be given below.

The correlation function can be related to single-particle distribution:
\begin{eqnarray}
\label{eq:2}
C(\eta_1,\eta_2) &=& \left\langle R(\eta_1)R(\eta_2)\right\rangle- \frac{\delta(\eta_1-\eta_2)}{\lr{N(\eta_1)}}, R(\eta)\equiv \frac{ N(\eta)}{\left\langle N(\eta)\right\rangle}
\end{eqnarray}
where $R(\eta)$ is the observed multiplicity density distribution in one event normalized by the ensemble average. In the absence of EbyE fluctuations, $R(\eta)=1$ and $C=1$.

One key step in our method is to decompose $R(\eta)$ into orthogonal polynomials in the rapidity range [-$Y$,$Y$]: 
\begin{eqnarray}
\label{eq:3}
R(\eta) = 1+\sum_n^{\infty}a_n\su{obs} \; T_n(\eta),\;\;\; T_n(\eta) \equiv \sqrt{n+\frac{1}{2}}P_n(\eta/Y)
\end{eqnarray}
where the $P_0(x)=1$, $P_1(x)=x$, $P_2(x)=1/2(3x^2-1)$..., are Legendre polynomials, and $Y$ characterizes the range of the rapidity fluctuations and is chosen to be $Y=6$ in current study. The superscript ``obs'' is used to explicitly denote the observed quantity in a single event. The new bases $T_n(x)$ are chosen such that their orthogonality and completeness relations are normalized as:
\begin{eqnarray}
\label{eq:4}
 1/Y\int_{-Y}^{Y} T_n(\eta)T_m(\eta) d\eta = \delta_{nm},\;\;  1/Y\sum_{n=0}^{\infty} T_n(\eta_1)T_n(\eta_2)=\delta(\eta_1-\eta_2)
\end{eqnarray}
Our approach is similar to that of Ref.~\cite{Bzdak:2012tp} except for two differences: 1) the decomposition is performed on deviation from average profile obtained in narrow centrality interval, instead of obtaining $\lr{N(\eta)}$ by averging over events with different $a_n$ values, and 2) the orthogonal bases are Legendre instead of Chebychev polynomials, the latter has a weight factor of $1/\sqrt{(1-(\eta/Y)^2)}$ in the normalization relation that diverges at $\eta=\pm Y$.

The $R(\eta)$ observable provides a natural way to separate the centrality dependence of the $\lr{N(\eta)}$ from the dynamical shape fluctuations for events within fixed centrality: the probability distribution of the $N(\eta)$ of all events, $p\{N(\eta)\}$, can be expressed as the sum of the product of the average shape $\lr{N(\eta)}_k$ and the probability distribution of multiplicity shape $p\{R(\eta)_k\}$ for centrality class ``$k$'':
\begin{eqnarray}
\label{eq:5}
p\{N(\eta)\} = \Sigma_k \lr{N(\eta)}_k p\{R(\eta)_k\}.
\end{eqnarray}
Events are first divided into narrow centrality classes according to their total multiplicity $M$ in $|\eta|<Y$. Next, the average multiplicity distribution $\left\langle N(\eta)\right\rangle$ is calculated for each event class, which is then used to calculate the EbyE $R(\eta)$. The coefficients of $T_n$ and their statistical uncertainty are calculated as:
 \begin{eqnarray}
\label{eq:6}
a_n\su{obs} = \Sigma_{i} w^n_i-\delta_{n}, \delta a_n\su{obs} = \sqrt{\Sigma_{i} \left(w^n_i\right)^2}, w^n_i = \frac{T_n(\eta_i)}{\lr{N(\eta_i)}}
\end{eqnarray}
where the sum is over all particles in the event, and $\delta_{n}=1$ for $n=0$ and 0 otherwise. The $\delta a_n\su{obs}$ characterizes the statistical fluctuations due to finite number of particles in the events, and so in principle it can be used to unfold the statistical smearing effects in $a_n\su{obs}$. In this paper, however, a more robust data-driven method is used to account for the smearing of $a_n\su{obs}$ due to finite number effect: for each real event, a random event is generated with same $M$ by sampling the $\left\langle N(\eta)\right\rangle$ and its coefficients $a_n\su{ran}$ are calculated using Eq.~\ref{eq:5}, which contain only the statistical effects. This method provides a simple but self-consistent treatment of the experimental effects.

Note that, the $T_n(\eta)$ bases are oscillating functions in pseudorapidity, in a way similar to the azimuthal flow harmonics, hence they are referred to as longitudinal harmonics. The non-statistical component of these longitudinal harmonics can be obtained after averaging over many events as:
\begin{eqnarray}
\label{eq:6a}
\lr{a_na_m} = \lr{a\su{obs}_na\su{obs}_m}-\lr{a\su{ran}_na\su{ran}_m}
\end{eqnarray}
A special case is the diagonal terms:
\begin{eqnarray}
\label{eq:6b}
\lr{a_n^2} = \lr{\lrp{a\su{obs}_n}^2}-\lr{\lrp{a\su{ran}_n}^2}.
\end{eqnarray}

The $a_n$ coefficients can also be obtained from the two-particle correlation function:
\begin{eqnarray}\nonumber
C(\eta_1,\eta_2) &=& 1+ \lr{R((\eta_1)R((\eta_2)}-\lr{R\su{ran}(\eta_1)R\su{ran}(\eta_2)} \\\nonumber
&=&1+ \sum_{n,m=0}^{\infty} \lrp{\lr{a_n\su{obs}a_m\su{obs}}-\lr{a_n\su{ran}a_m\su{ran}}}T_n(\eta_1)T_n(\eta_2)\\\nonumber
&=&1+ \sum_{n,m=0}^{\infty} \lr{a_na_m}T_n(\eta_1)T_m(\eta_2)\\
&=&1+\sum_{n,m=0}^{\infty} \lr{a_na_m}\frac{T_n(\eta_1)T_m(\eta_2)+T_n(\eta_2)T_m(\eta_1)}{2}\label{eq:7a}
\end{eqnarray}
where we have used the fact that the $R\su{ran}(\eta_1)$ and $R\su{ran}(\eta_2)$ are uncorrelated except at $\eta_1=\eta_2$. In other words, one could construct a correlation function from random events, then it can be shown that:
\begin{eqnarray}
C\su{ran}(\eta_1,\eta_2)\equiv\lr{R\su{ran}(\eta_1)R\su{ran}(\eta_2)}=1+\frac{\delta(\eta_1-\eta_2)}{\left\langle N(\eta_1)\right\rangle}
\end{eqnarray}
This means that the correlation function excluding self-pairs gives directly the $\lr{a_na_m}$ as the statistical effects drop out after averaging pairs over many events. The last part of Eq.~\ref{eq:7a} is required by $C(\eta_1,\eta_2)=C(\eta_2,\eta_1)$. Furthermore, symmetric collision systems such as Pb+Pb require $C(\eta_1,\eta_2)~=~C(-\eta_1,-\eta_2)$, leading to $\lr{a_na_{n+1}}=0$, i.~e. odd and even harmonics are uncorrelated. The remaining coefficients can be calculated analytically from the correlation function as:
\begin{eqnarray}\label{eq:7b}
\lr{a_na_m} =\frac{1}{Y^2} \int \left[C(\eta_1,\eta_2)-1\right] \frac{T_n(\eta_1)T_m(\eta_2)+T_n(\eta_2)T_m(\eta_1)}{2} d\eta_1d\eta_2
\end{eqnarray}

\begin{figure}[!t]
\begin{center}
\includegraphics[width=1\columnwidth]{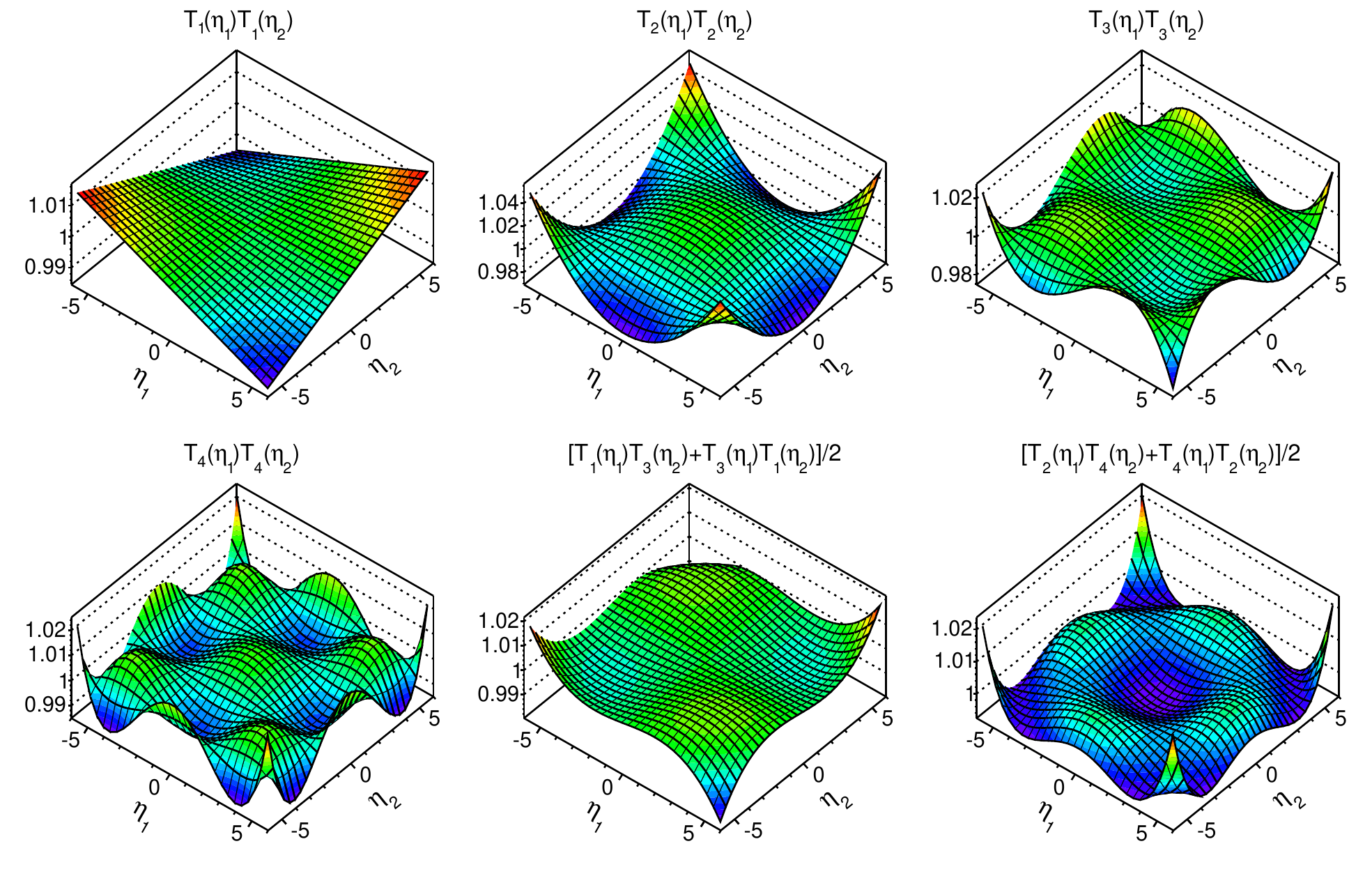}
\end{center}
\caption{\label{fig:0}  The shape of the first few bases associated with $\lr{a_na_m}$ in the two-particle correlation function. They are plotted assuming $\lr{a_na_m}=0.01$.}
\end{figure}

Figure~\ref{fig:0} shows the expected shape of the bases in the correlations function, they are plotted assuming $\lr{a_na_m}=0.01$. The base for the first term $\lr{a_1a_1}$ is proportional to $\eta_1\eta_2$ and is characterized by quadratic shape along $\eta_1=\eta_2$ and $\eta_1=-\eta_2$ but with opposite sign (see similar discussion in Ref.~\cite{Bzdak:2012tp}). The base for $\lr{a_2a_2}$ is characterized by four sharp peaks at the four corners of the correlations function and a broader peak around $\eta_1=\eta_2\approx0$.

The single-particle method denoted by Eqs.~\ref{eq:3} and \ref{eq:6a} and the correlation method denoted by Eqs.~\ref{eq:7a} and ~\ref{eq:7b} are mathematically equivalent. The single-particle method calculates $a_n\su{obs}$ for each event and hence allows direct correlation with its initial geometry in model calculations. Furthermore, it also allow study of possible non-Gaussianity in the distribution of $a_n$. On the other hand, the correlation method calculates all $\lr{a_na_m}$ in a single pass, and systematic effects from experiments are easier to control (e.g. via mixed events).

The discussion above can be generalized into correlations of more than three coefficients, such as $\lr{a_na_ma_l}$. For the single-particle method, it just requires a simple extension of the Eq.~\ref{eq:6a}; while multi-particle correlation functions are required for the correlation method, e.~g. $C(\eta_1,\eta_2,\eta_3)$~\footnote{The multi-particle correlation function is closely related to the multi-bin correlator proposed in Ref.~\cite{Bialas:2011xk}}. This is an interesting avenue that deserves further studies.

To demonstrate the robustness and physics potential of the method, we carried out a detailed simulation study using the HIJING~\cite{Gyulassy:1994ew} and AMPT~\cite{Lin:2004en} models. The HIJING model combines the lund-string dynamics for soft particle production and hard QCD interaction for high $\pT$ particle production, which naturally contains many sources of long-range and short-range correlations. The AMPT model starts from the particles produced by HIJING, breaks them into partons (``string-melting'') and runs them though partonic transport. The partons are then recombined to form hadrons at freezeout density, which in turn undergo hadronic transport. The partonic transport processes generate significant collective flow and was demonstrated to qualitatively describe the harmonic flow $v_n$ in $p$+A and A+A collisions~\footnote{The model simulation is performed with the string-melting mode with a total partonic cross-section of 1.5 mb and strong coupling constant of $\alpha_s$ = 0.33. This setup has been shown to reproduce the experimental $\pT$ spectra and $v_n$ data at RHIC and the LHC.}. Therefore measuring the longitudinal harmonics $a_n$ in HIJING and AMPT models allows us to understand how longitudinal multiplicity fluctuations in the early time are affected by the final-state interactions.

The HIJING and AMPT data used in this study are generated for Pb+Pb collisions at LHC energy of $\sqrt{s_{\mathrm{NN}}}=2.76$ TeV. All stable particles with $\pT>0.1$ GeV/$c$ in the pseudorapidity range of $|\eta|<Y=6$ are used. In the default setup, events are first sorted into narrow event activity classes based on total multiplicity $M$, i.~e. the $M$ of all events in each class is required to differ from the average multiplicity of event class by at most 1\%. The $N(\eta)$ distribution is then obtained for each event and the $a_n\su{obs}$ coefficients are calculated. At the same time, a random event containing $M$ particles is generated according to $\lr{ N(\eta)}$ and the coefficients $a_n\su{ran}$ are obtained. The same classification is also used for 2PC method, however the $\lr{a_na_m}$ are calculated directly via Eq.~\ref{eq:7b} without using the random events. This event classification procedure in obtaining $\lr{ N(\eta)}$ allows a separation of the centrality dependence of the shape of the $N(\eta)$ distribution (controlled by $M$) from the shape fluctuations for events with the same $M$. Hence we can get a clearer understanding of the dynamic FB multiplicity fluctuations separated from the overall multiplicity fluctuation. For comparison, $\lr{ N(\eta)}$ is also obtained using event classes based on either $\npart$ or impact parameter $b$, where much stronger EbyE fluctuation is expected for $R(\eta)$.

In the following, we discuss the properties of the $a_n$ coefficients based on results obtained from the single-particle method. However, most of these results can be also obtained with the 2PC method. 

\section{Properties of longitudinal harmonics from the single-particle method}
\label{sec:sp}
\begin{figure}[!t]
\begin{center}
\includegraphics[width=1\columnwidth]{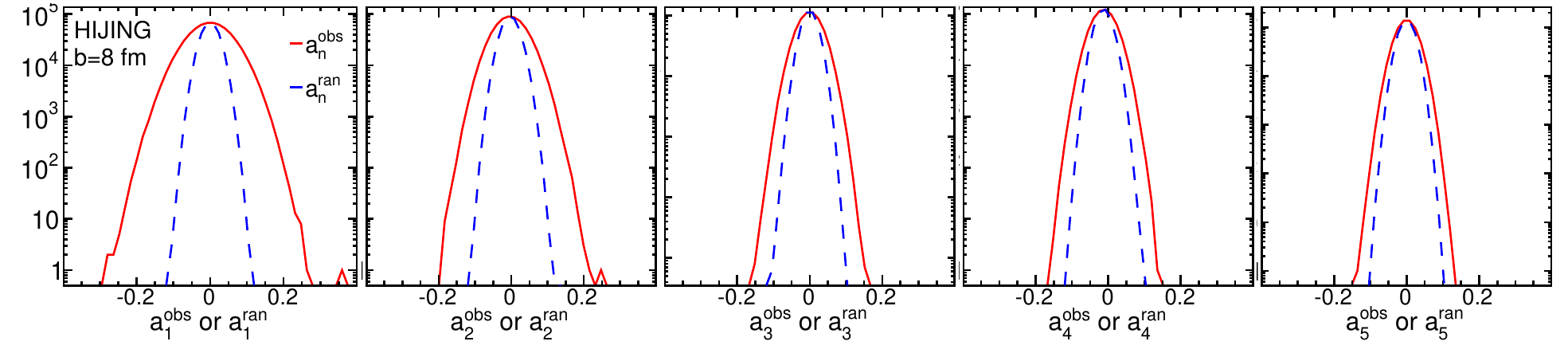}
\includegraphics[width=\columnwidth]{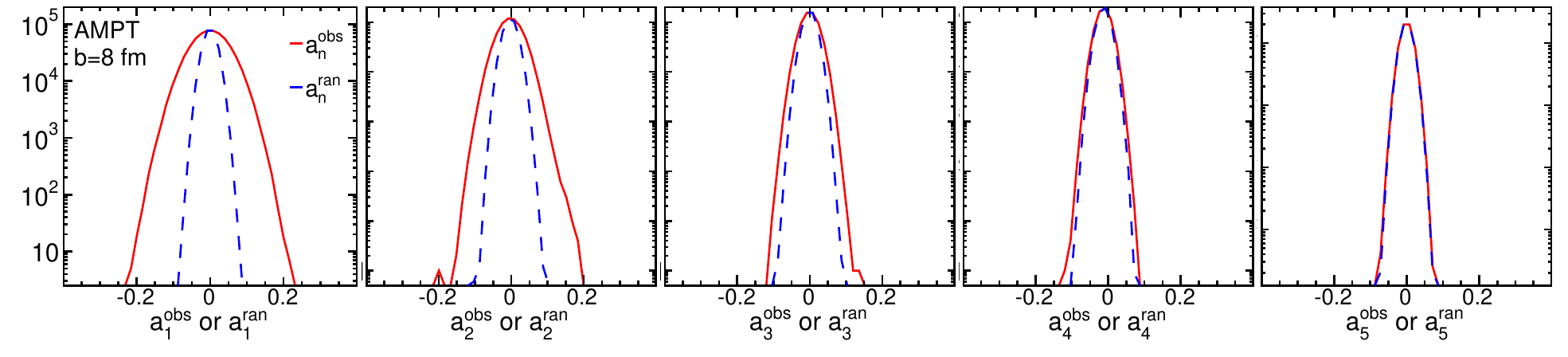}
\end{center}
\caption{\label{fig:1}  The distributions of coefficients for longitudinal Legendre polynomials from real events $a_n\su{obs}$ and random events $a_n\su{ran}$ for HIJING (top row) and AMPT (bottom row) events with $b=8$~fm. The panels in each row correspond to results from $n=1$ to $n=5$.}
\end{figure}

\begin{figure}[!t]
\begin{center}
\includegraphics[width=0.45\columnwidth]{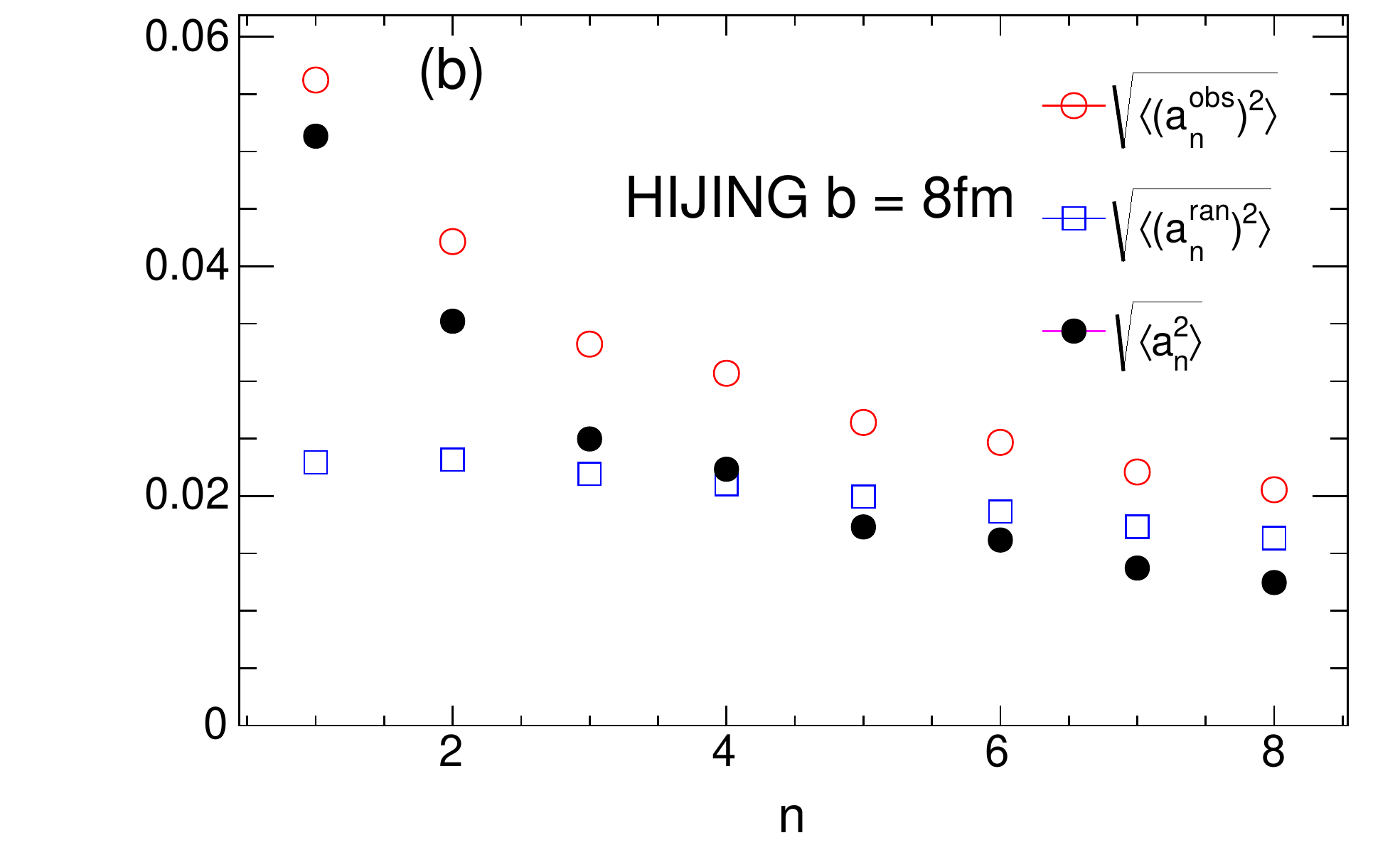}\includegraphics[width=0.45\columnwidth]{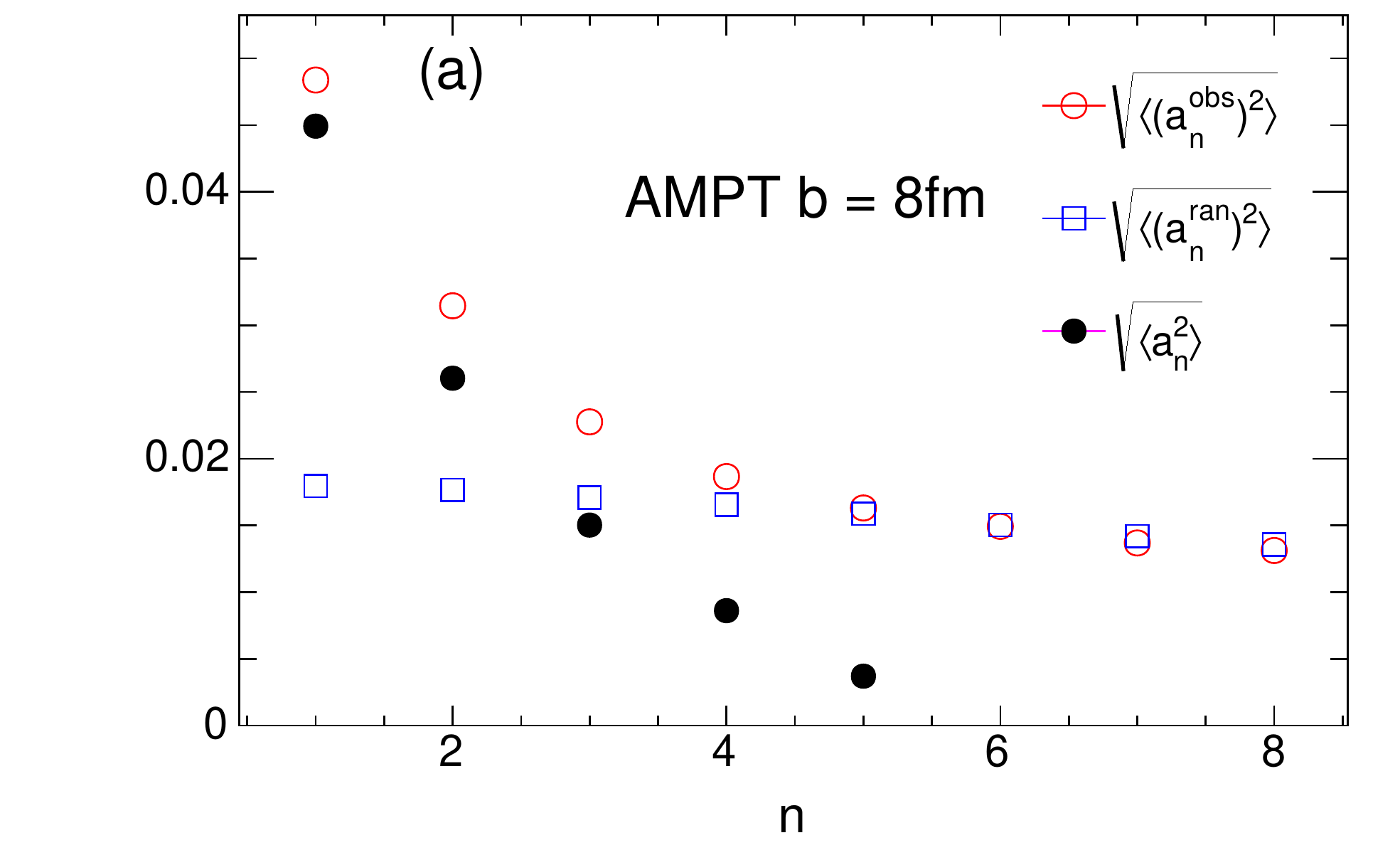}
\end{center}
\caption{\label{fig:2a}  The $a_n$ vs $n$ from HIJING (left) and AMPT(right) events with $b=8$~fm.}
\end{figure}

\begin{figure}[!t]
\begin{center}
\includegraphics[width=1\columnwidth]{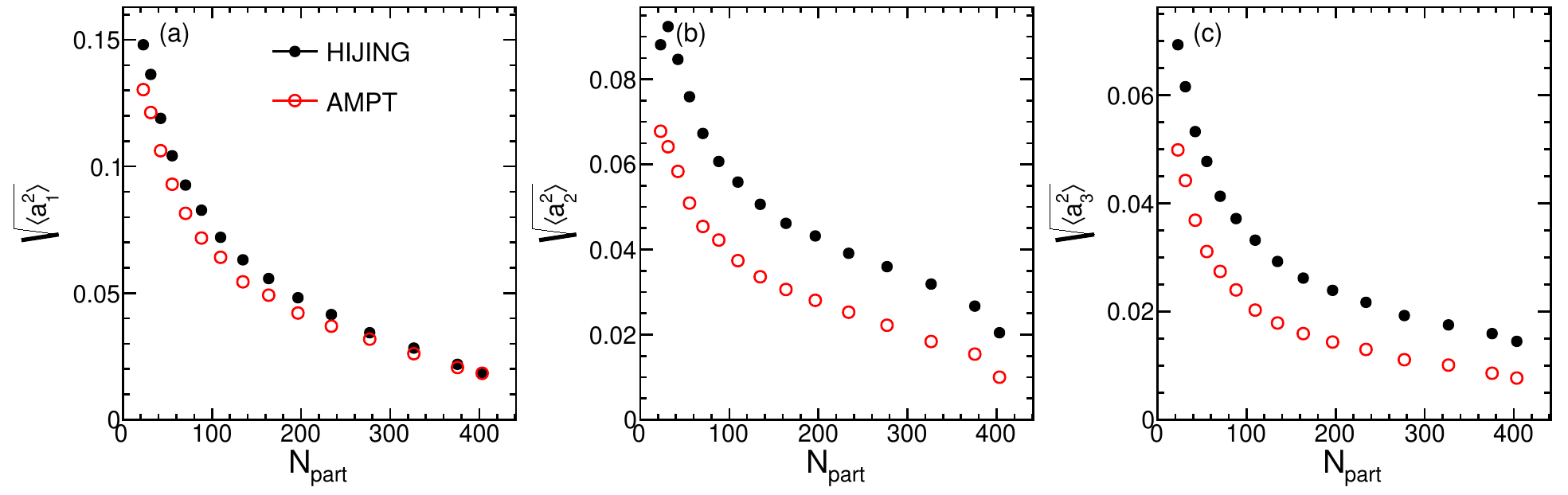}
\end{center}
\caption{\label{fig:2b}  Centrality dependence of $a_1$ (left panel), $a_2$ (middle panel) and $a_3$ (right panel) for HIJING and AMPT events.}
\end{figure}

Figure~\ref{fig:1} shows the EbyE distributions of $a_n\su{obs}$ for events with fixed impact parameter $b=8$~fm, and they are compared with distributions obtained from random events $a_n\su{ran}$. The differences between the two types distributions reflect dynamical fluctuations in $a_n\su{obs}$. These differences decrease for larger $n$, and the rate of decrease is much larger in AMPT events than in HIJING events. By $n=5$, the distribution for AMPT events is consistent with pure statistical fluctuation. From these distributions, the $\lr{a_n^2}$ signals are extracted via Eq.~\ref{eq:6b} and shown as a function of $n$ in Fig.~\ref{fig:2a}. Significant values of $a_n$ are seen for all harmonics in HIJING events, while they decrease rapidly and are consistent with zero for $n>4$ in AMPT events. This difference is mainly due to stronger short-range correlations present in HIJING events (see Fig.~\ref{fig:7c}), but could also due to strong viscous damping associated with final-state rescatterings in the AMPT model. Figure~\ref{fig:2b} compares the centrality dependence of the $a_1$, $a_2$ and $a_3$ in HIJING and AMPT models. The signal strength increases towards more peripheral collisions and the values from AMPT model are consistently smaller than those from HIJING in all centrality ranges.

\begin{figure}[!t]
\begin{center}
\includegraphics[width=1\columnwidth]{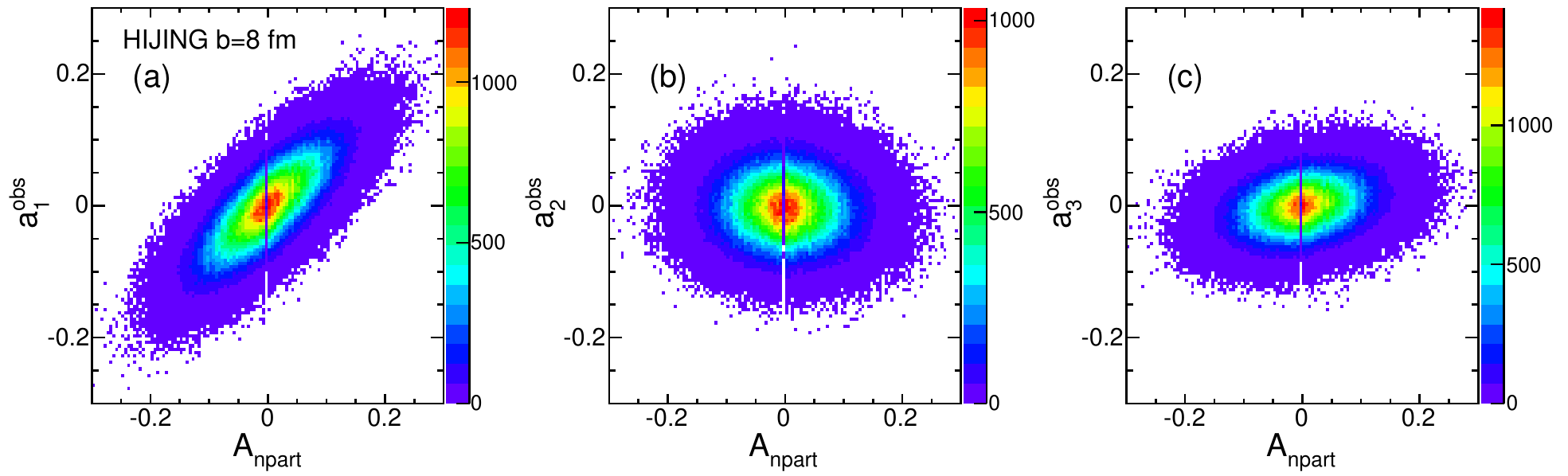}
\end{center}
\caption{\label{fig:3}  Event-by-event correlation between $a_n\su{obs}$ and $A\sub{part}$ for $n=1$ (left panel), $n=2$ (middle panel) and $n=3$ (right panel) from HIJING events with $b=8$ fm.}
\end{figure}

In order to find out whether the FB multiplicity fluctuation is related to the difference between $\npartf$ and $\npartb$, $a_n\su{obs}$ is correlated directly with $A\sub{part}$, defined as:
\begin{eqnarray}
\label{eq:10}
A\sub{part}=\frac{\npartf-\npartb}{\npartf+\npartb}\;.
\end{eqnarray}
The results for $b=8$~fm from HIJING events are shown in Fig.~\ref{fig:3} (results for AMPT events are similar). A strong positive correlation between $a_1\su{obs}$ and $A\sub{part}$ is observed, suggesting that the FB asymmetry in the multiplicity distribution is indeed driven by the asymmetry in the number of participating nucleons in the two colliding nuclei. A weak correlation is also observed between $a_3\su{obs}$ and $A\sub{part}$, suggesting that the FB asymmetry caused by $A\sub{part}$ contains a small non-linear odd component. On the other hand, there is no correlation between $a_2\su{obs}$ (rapidity even) and $A\sub{part}$ (rapidity odd) as expected. The width of these distributions are partially due to statistical smearing effects in $a_n\su{obs}$, which can be removed by a 2D unfolding (leave for a future work).

\begin{figure}[!t]
\begin{center}
\includegraphics[width=0.8\columnwidth]{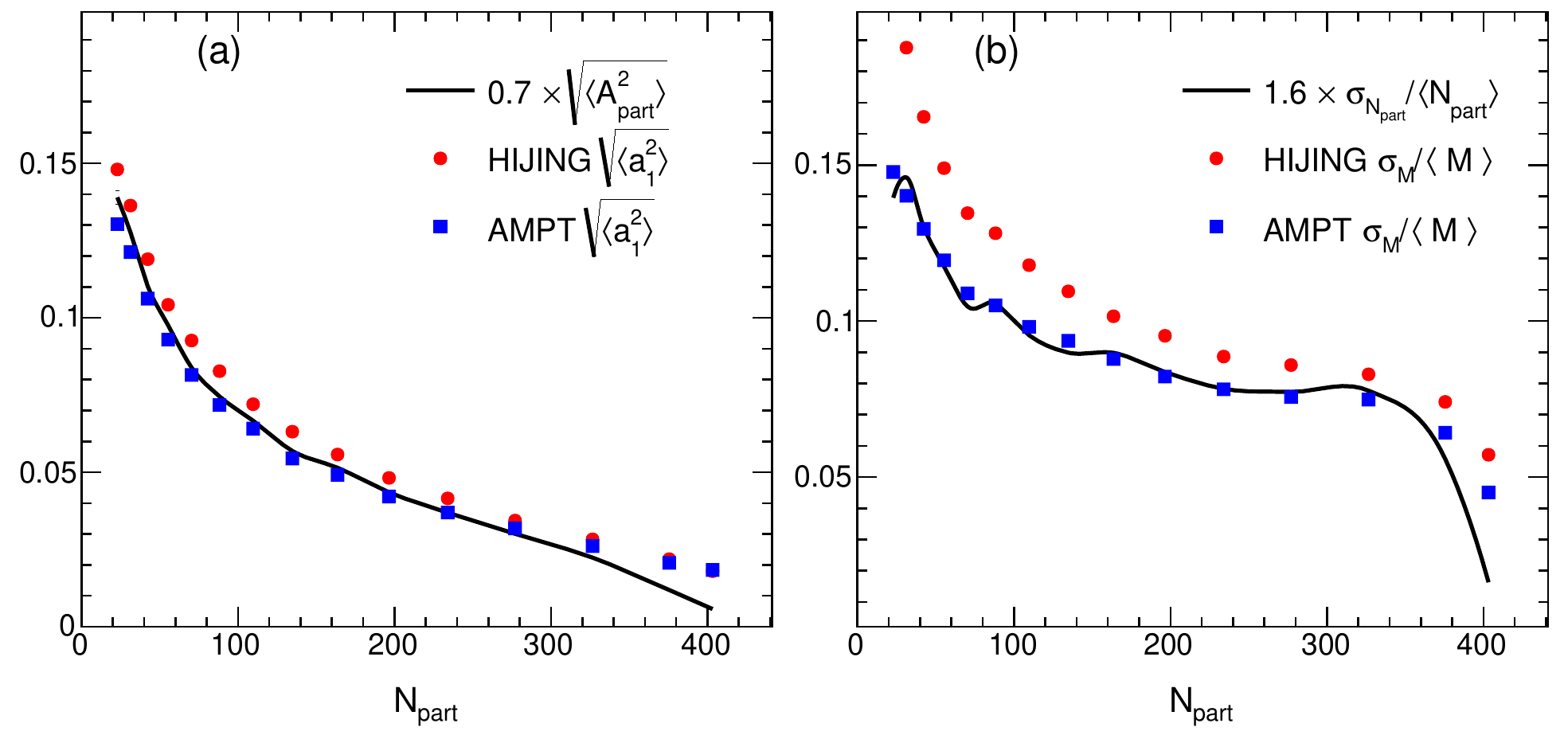}
\end{center}
\caption{\label{fig:4}  Comparison between $\sqrt{\lr{a_{1}^2}}$ and RMS asymmetry in $N\sub{part}$, $\sqrt{\lr{A\sub{part}^2}}$ (left panel), as well as between total multiplicity fluctuation in terms of $\sigma_{N_{ch}}/\lr{N_{ch}}$ and fluctuation of total $N\sub{part}$ (right panel) in HIJING and AMPT models.}
\end{figure}

Figure~\ref{fig:4} (a) compares the centrality dependence of $\sqrt{\lr{a_1^2}}$ and $\sqrt{\lr{A\sub{part}^2}}$. The similarity in their shapes suggest that the asymmetry between $\npartf$ and $\npartb$ is primarily responsible for the FB asymmetry in $N(\eta)$. Note that the FB asymmetry of $R(\eta)$ arising from $a_1$ can be estimated as, $A_R(\eta) \approx \sqrt{\lr{a_1^2}}T_1(\eta) = \sqrt{\frac{3}{2}} \sqrt{\lr{a_1^2}}\frac{\eta}{6}$. The results in Fig.~\ref{fig:4} (a) imply $\sqrt{\lr{a_1^2}} \approx  0.7 \sqrt{\lr{A\sub{part}^2}}$, and hence $A_R(6)=\sqrt{\frac{3}{2}} \sqrt{\lr{a_1^2}} \approx 0.86 \sqrt{\lr{A\sub{part}^2}}$. Therefore, the multiplicity fluctuations in the very forward (backward) rapidity ($\pm6$) are mostly driven by the fluctuations in $\npartf$ ($\npartb$). On the other hand, the fluctuation of total multiplicity $M$ is expected to be driven mainly by the fluctuation of $\npart=\npartf+\npartb$. Given that $a_1$ is driven by $\npartf-\npartb$, the fluctuation of $M$ should not be independent from fluctuation of $a_1$. Figure~\ref{fig:4} (b) compares the relative multiplicity fluctuation, $\sigma_M/\lr{M}$, with the fluctuation of number of participants $\sigma_{\npart}/\lr{\npart}$. Indeed, the two show very similar centrality dependence after applying a constant scale factor. 

\begin{figure}[!t]
\begin{center}
\includegraphics[width=1\columnwidth]{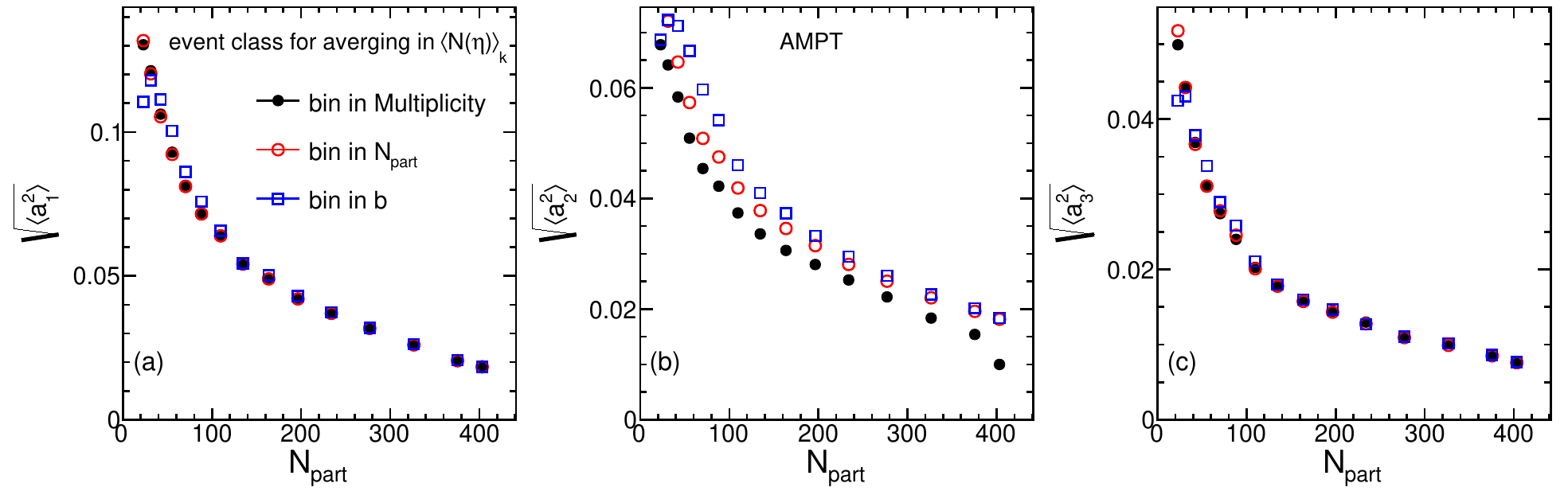}
\end{center}
\caption{\label{fig:5}  Comparison of the $a_n$ obtained from three averaging methods, i.e. binning in total multiplicity, $\npart$ or impact parameter $b$, for $\lr{ N(\eta)}$ used in Eq.~\ref{eq:2} for $n=1$ (left panel), $n=2$ (middle panel) and $n=3$ (right panel).}
\end{figure}

\begin{figure}[!h]
\begin{center}
\includegraphics[width=0.8\columnwidth]{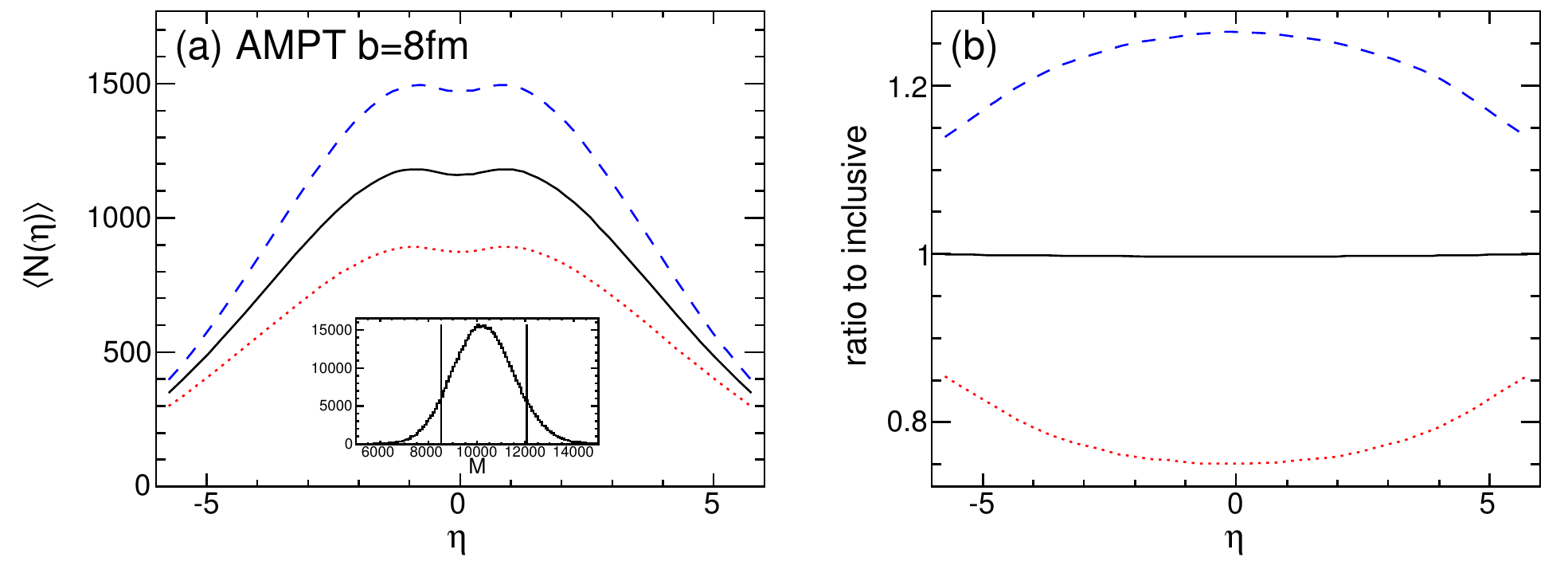}
\end{center}
\caption{\label{fig:6}  (a) The average multiplicity distributions for events selected in three multiplicity ranges (see insert) and (b) the ratios to the all events. All events are generated for AMPT model with $b=8$ fm.}
\end{figure}

The results shown so far are obtained by calculating $\lr{ N(\eta)}$ in narrow bins of $M$. Figure~\ref{fig:5} compare these with results obtained in narrow slices of $\npart$ or $b$. This comparison is useful because experiments can only measure $\lr{ N(\eta)}$ in finite centrality interval for which the overall multiplicity can still have significant fluctuations. Figure~\ref{fig:5} shows that the values of $a_1$ and $a_3$ have very weak dependence on the averaging scheme, while $a_2$ has rather strong dependence. The latter suggests that a significant component of the $a_2$ obtained for binning in $\npart$ or $b$ arises from the residual centrality dependence in the shape of $\lr{ N(\eta)}$. To see how this residual centrality dependence can arise, Fig.~\ref{fig:6} compares the $\lr{ N(\eta)}$ obtained for events in the upper or lower tails of the total multiplicity distribution for all events with $b=8$~fm. The ratios on the right panel show that the shape of $\lr{ N(\eta)}$ can still vary significantly for events with the same impact parameter but different $M$, and this variation leads to a significant $a_2$ contribution. Nevertheless, after removing this residual centrality dependence by binning events in narrow $M$ ranges, a significant $a_2$ signal still remains. This irreducible $a_2$ could reflect strong event-by-event fluctuations in the amount of nuclear stopping or shift of the effective center-of-mass of the collisions~\cite{Steinberg:2007fg,Vovchenko:2013viu}. Similar results are also seen in HIJING events (not shown). 

\section{Correlating $a_1$ with spectator asymmetry}
\label{sec:spec}
If the $a_1$ coefficient is correlated with the fluctuations of $\npartf-\npartb$, then it should be anti-correlated with the asymmetry in the number of spectator nucleons $\nspecf-\nspecb$ since:
\begin{eqnarray}
\label{eq:11}
\npartf-\npartb=-(\nspecf-\nspecb).
\end{eqnarray}
The number of spectator nucleons can be measured using calorimeters placed very close to the beam-line in the forward region. For example, the Zero-degree Calorimeters (ZDC) installed in all RHIC and LHC experiments can count the number of spectator neutrons, $\nneu$, in each event with rather good precision. Unfortunately, the measured neutrons only constitute a small fraction of all spectator nucleons, and hence the correlation between $\npartf-\npartb$ and FB neutron asymmetry $\nneuf-\nneub$ is expected to be very weak. Nevertheless, studying the correlation between $a_1$ and $\nspecf-\nspecb$ provides an independent and data-driven way for understanding the origin of the FB multiplicity correlations. 

\begin{figure}[!t]
\begin{center}
\includegraphics[width=1\columnwidth]{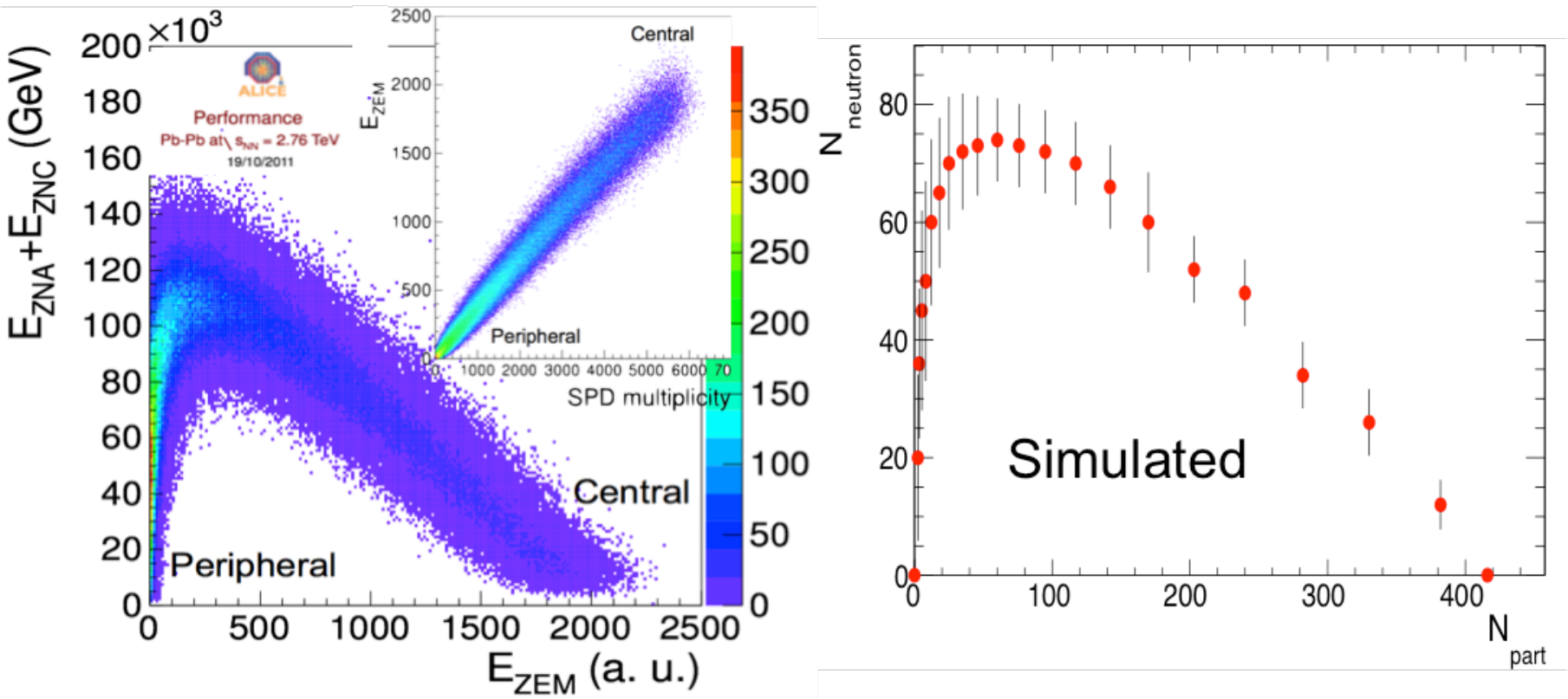}
\end{center}
\caption{\label{fig:7}  a) The correlation of signals in ZDC and ZEM from ALICE experiment, the insert shows the correlation of signals in ZEM and SPD. Then number of neutrons are calculated as $\nneu = E_{\mathrm{ZDC}}/1.38$. b) The inferred correlation between $\nneu$ and $\npart$ used in this paper.}
\end{figure}

Figure~\ref{fig:7}(a) shows the ALICE measurement of the correlation of the ZDC energy with ZEM ($4.8<|\eta|<5.7$) energy in Pb+Pb collisions at $\sqrt{s_{\rm{NN}}}=2.76$ TeV~\cite{Abelev:2013qoq}. The latter has a very strong correlation with the Silicon Pixel Detector (SPD) situated in mid-rapidity ($|\eta|<1.9$) as shown by the insert panel. The ZEM signal can be mapped onto the $\npart$ assuming $E_{\mathrm{ZEM}}\propto \npart$, and the ZDC signal is converted to $\nneu$ from the expected energy for each spectator nucleon of 1.38 TeV:  $\nneu = E_{\mathrm{ZDC}}/1.38$. From this, the correlation between $\npart$ and the average number of neutrons $\lr{\nneu}$ is estimated and shown in Fig.~\ref{fig:7}(b), where the error bars indicate the approximate standard deviations. This correlation is then down-scaled by a factor of two in both axes to give the correlation between $\npartf$ and $\lr{\nneuf}$ or between $\npartb$ and $\lr{\nneub}$. However, the error bar is reduced only by a factor of $\sqrt{2}$ assuming the sampling of $\nneuf$ is independent of $\nneub$ once the values of $\npartf$ and $\npartb$ are fixed in each event (hence $\nspecf=208-\npartf$ and $\nspecb=208-\npartb$ are also fixed). This new distribution is then used to generate the $\nneuf$ and $\nneub$ for each HIJING or AMPT event based on its $\npartf$ and $\npartb$ values. Finally we calculate the correlation between $\nneuf-\nneub$ and $a_1\su{obs}$.

\begin{figure}[!t]
\begin{center}
\includegraphics[width=1\columnwidth]{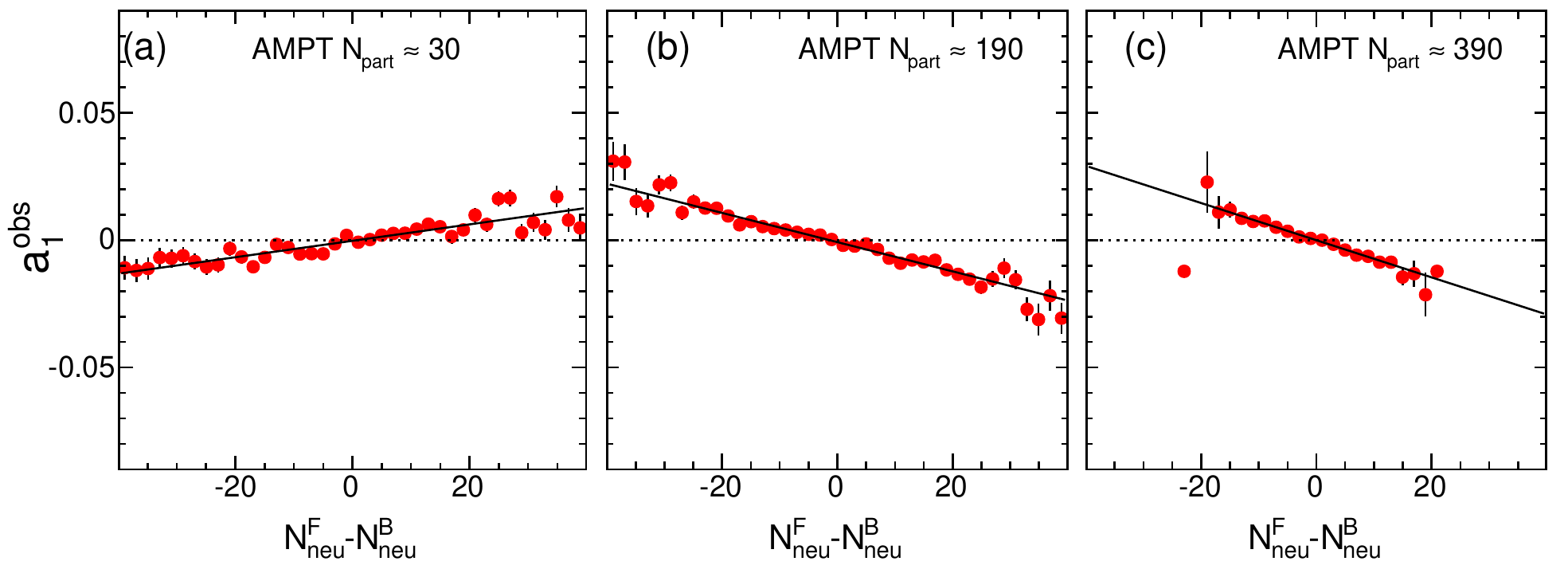}
\end{center}
\caption{\label{fig:8}  The estimated correlation between $a_1\su{obs}$ and $\nneuf-\nneub$ for peripheral (left panel), mid-central (middle panel) and central (right panel) Pb+Pb collisions.}
\end{figure}

The results of this study for AMPT events is summarized in Fig.~\ref{fig:8}.  A clear anti-correlation is seen in mid-central and central collisions. However the correlation is positive in peripheral collisions, which reflects the fact that the value of $\nneu$ is positively correlated with $\npart$ in the peripheral collisions (see Fig.~\ref{fig:7}(a)). This correlation is very weak, $a_1\su{obs}$ varies by a few percent in the available range of $\nneuf-\nneub$, but should be measurable in experiments.

\section{Additional insights from two-particle correlation method}
\label{sec:2pc}

\begin{figure}[!t]
\begin{center}
\includegraphics[width=0.8\columnwidth]{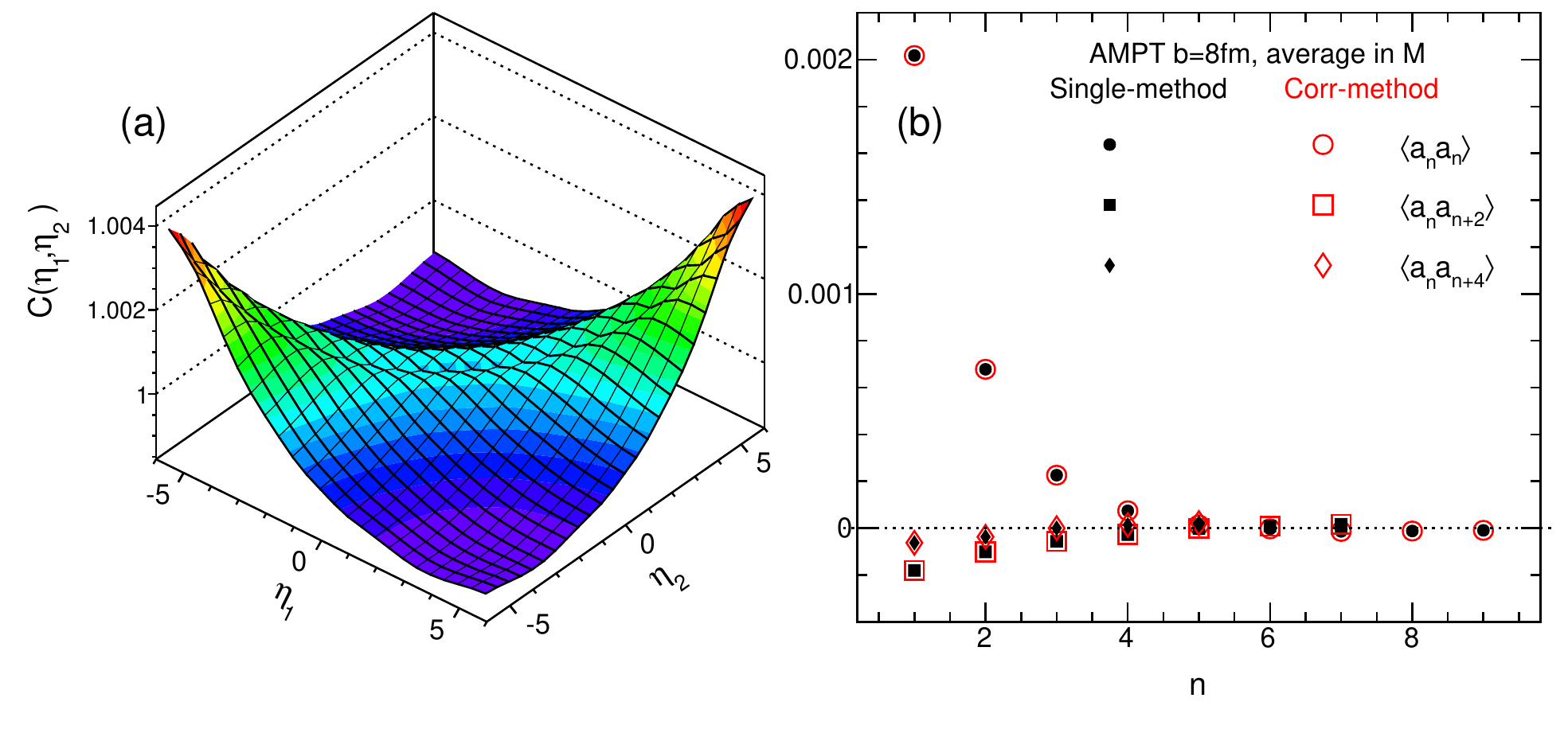}
\end{center}
\caption{\label{fig:7a}  The correlation function (left) and corresponding spectrum $\lr{a_na_m}$ for $n,m\le9$ (right panel) for AMPT events generated with $b=8$~fm, where the $\lr{N(\eta)}$ is calculated in narrow multiplicity bins.  The spectrum are compared with those calculated directly from the single-particle method.}
\end{figure}

As discussed in Sec.~\ref{sec:method}, $a_n$ coefficients can also be calculated from correlation method via Eq.~\ref{eq:7b}. Figure~\ref{fig:7a} (a) shows the correlation function and $\lr{a_na_m}$ values from AMPT events with $b=8$~fm. The shape of the correlation function already suggests the dominance of the $\lr{a_1^2}$ term (compare with Fig.~\ref{fig:0}). The coefficients are compared with those obtained from the single-particle method via Eq.~\ref{eq:6a}, identical values are observed. This consistency is expected since the two methods are mathematically equivalent. A selected set of coefficients are shown in Fig.~\ref{fig:7a} (b). No correlations are observed between the odd and even coefficients as expected for symmetric collision system, while small anti-correlations are observed between odd or even terms, i.e. $\lr{a_na_{n+2}}<0$ and $\lr{a_na_{n+4}}<0$.

One important practical advantage of the 2PC method is that it provides a natural way to separate the residual centrality dependence of average shape of $N(\eta)$ from the dynamical shape fluctuations for events with the same centrality. Eq.~\ref{eq:7a} can be rewritten as :
\begin{eqnarray}\label{eq:12}
C(\eta_1,\eta_2)= 1+\frac{1}{2}\lr{a_0a_0}+\frac{1}{\sqrt{2}}\sum_{n=1}^{\infty} \lr{a_0a_n}(T_n(\eta_2)+T_n(\eta_1))+ \sum_{n,m=1}^{\infty} \lr{a_na_m}\frac{T_n(\eta_1)T_m(\eta_2)+T_n(\eta_2)T_m(\eta_1)}{2}
\end{eqnarray}
The first term $ \lr{a_0a_0}$ reflects the multiplicity fluctuation in the given event class, which drops out from the expression if $C(\eta_1,\eta_2)$ is normalized to have a mean value of one (we shall assume that in the following discussion). The second term represents residual centrality dependence in the shape of $\lr{N(\eta)}$. The last term encodes the dynamical shape fluctuations for events with fixed centrality, which can be isolated by dividing the correlation function by its projections on the $\eta_1$ and $\eta_2$ axes:
\begin{eqnarray}
\label{eq:13}
&&C\sub{N}(\eta_1,\eta_2)= \frac{C(\eta_1,\eta_2)}{C_p(\eta_1)C_p(\eta_2)}\\
&&C_p(\eta_1) = \frac{\int C(\eta_1,\eta_2) d\eta_2}{2Y},C_p(\eta_2) = \frac{\int C(\eta_1,\eta_2) d\eta_1}{2Y}.
\end{eqnarray}
The new correlation function ensures that any residual centrality dependence is taken out from the measured coefficients:
\begin{eqnarray}\label{eq:14}
C_{\rm{N}}(\eta_1,\eta_2)= 1+\sum_{n,m=1}^{\infty} \lr{a^{'}_na^{'}_m}\frac{T_n(\eta_1)T_m(\eta_2)+T_n(\eta_2)T_m(\eta_1)}{2}
\end{eqnarray}
where the new coefficients are:
\begin{eqnarray}\label{eq:15}
\lr{a^{'}_na^{'}_m}\approx \lr{a_na_m}- \lr{a_0a_n}\lr{a_0a_m}\;.
\end{eqnarray}
They differ from the original coefficients by a small term $\lr{a_0a_n}\lr{a_0a_m}$, representing the contribution from the residual centrality dependence. Alternatively, $C_{N}$ can also be defined as:
\begin{eqnarray}
\label{eq:15b}
C\sub{N}(\eta_1,\eta_2)=  C(\eta_1,\eta_2)+1-C_p(\eta_1) C_p(\eta_2) 
\end{eqnarray}
or:
\begin{eqnarray}
\label{eq:15c}
C_{\rm{N}}(\eta_1,\eta_2)= C(\eta_1,\eta_2)+2-C_p(\eta_1)-C_p(\eta_2)\;.
\end{eqnarray} 
Equation~\ref{eq:15b} practically gives the same answer as Eq.~\ref{eq:13}. Eq.~\ref{eq:15c} is not preferred as it does not remove the $\lr{a_0a_n}\lr{a_0a_m}$ contribution in $\lr{a_na_m}$, although in practice the relative difference between the two is only a few percent. For all results shown below, definition Eq.~\ref{eq:13} is used.

\begin{figure}[!t]
\begin{center}
\includegraphics[width=1\columnwidth]{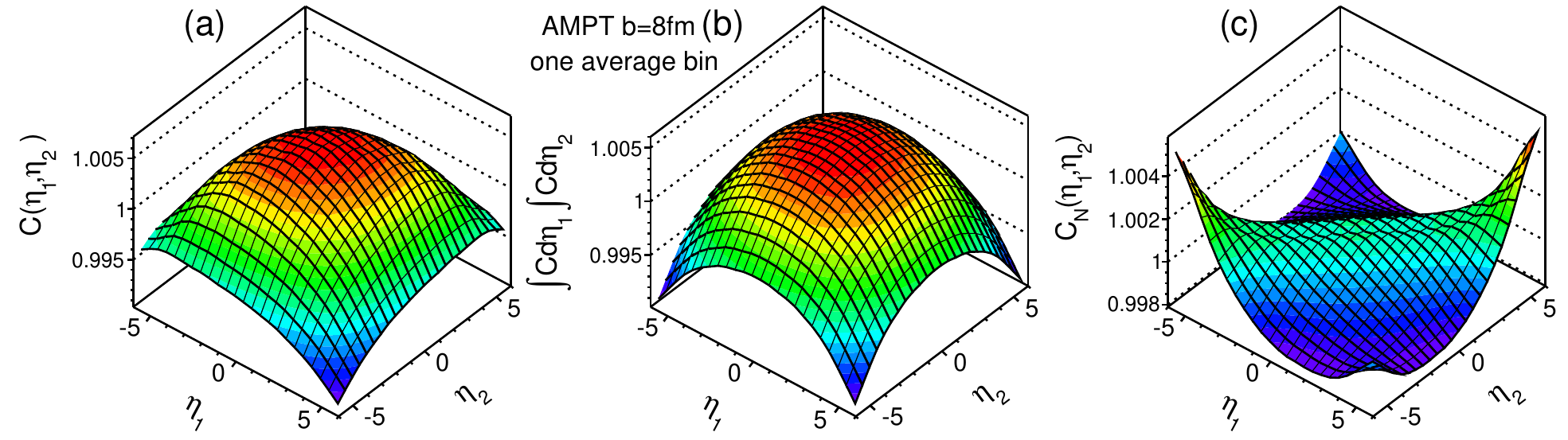}
\end{center}
\caption{\label{fig:7b}  The correlation function (left), the product of the projections on two axes (middle) and the redefined correlation function via Eq.~\ref{eq:13} (right panel) for AMPT events generated with $b=8$~fm. The $\lr{N(\eta)}$ is calculated using all events. The shallow dip structure shown in the right panel is already present in the left panel.}
\end{figure}

Figure~\ref{fig:7b} shows the original correlation function, the product of its projections to the two axes, and the renormalized correlation function for AMPT events for $b=8$~fm, where the average distribution $\lr{N(\eta)}$ is calculated in one bin (as appose to many narrow multiplicity bins then summed as in Fig.~\ref{fig:7a}). Despite the significant difference in the original correlation function due to the residual centrality dependence, the renormalized correlation function is very similar to that shown in Fig.~\ref{fig:7b}. The small difference in the four corners of the correlation functions can be attributed to the difference in $\lr{a_2^2}$ between different binning schemes shown in Fig.~\ref{fig:5}(b). Thus the $C_{\rm{N}}(\eta_1,\eta_2)$ defined in Eq.~\ref{eq:14} provides a robust way to extract the dynamical shape fluctuations nearly independent of the choice of centrality classes.

\begin{figure}[!t]
\begin{center}
\includegraphics[width=0.85\columnwidth]{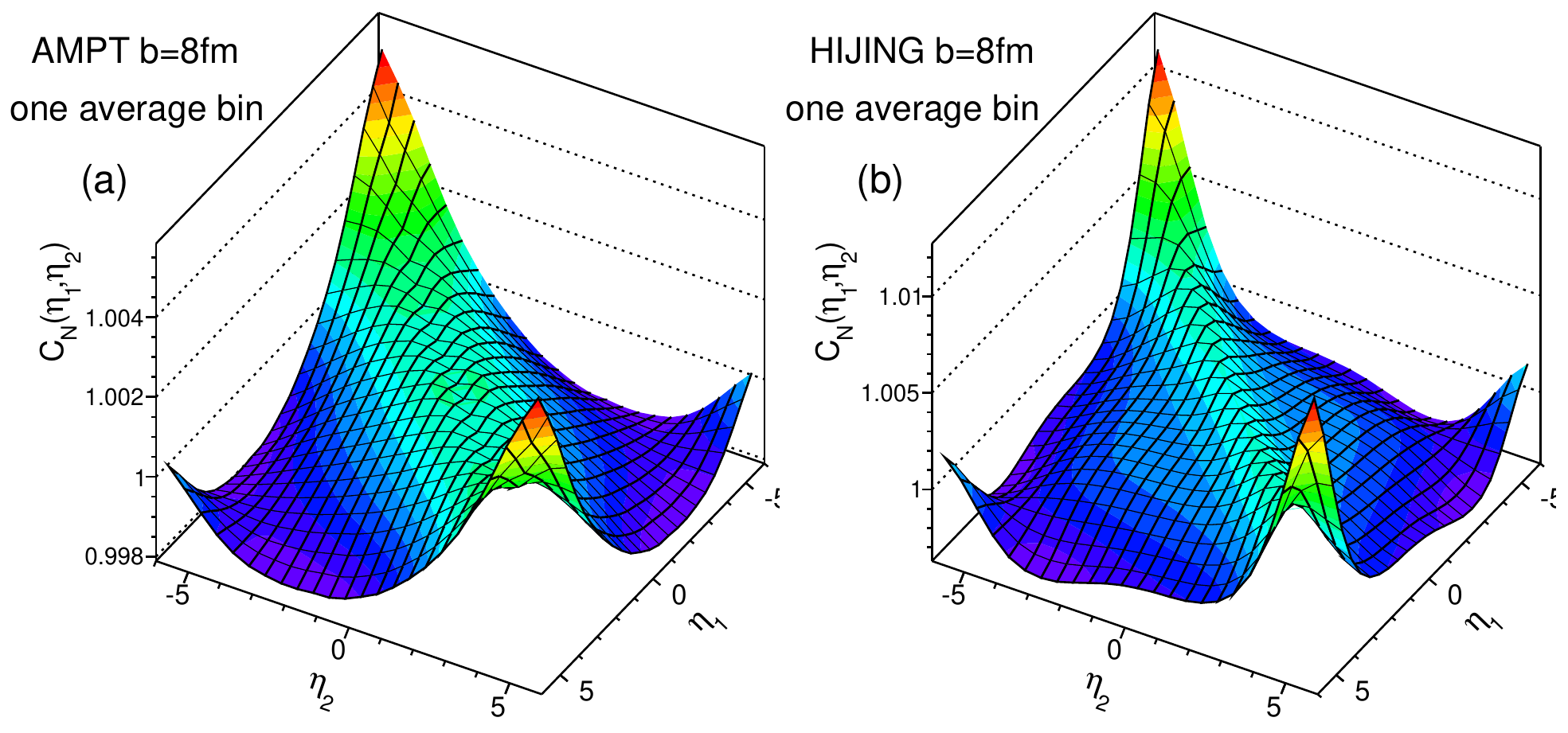}
\end{center}
\caption{\label{fig:7c}  The correlation function defined via Eq.~\ref{eq:13} for AMPT (left) and HIJING (right) events generated with $b=8$~fm. The $\lr{N(\eta)}$ is calculated using all events.}
\end{figure}

Figure~\ref{fig:7c} compares the correlation functions between the HIJING and AMPT, the correlation function from AMPT appears much broader than the HIJING, which is partially responsible for the faster decrease of the spectrum shown in Fig.~\ref{fig:2a}. The AMPT events also show an interesting shallow minimum around $\Delta\eta=0$ with a width of about $\pm0.4$. Since it is absent in HIJING events, this structure must reflect the influence of the final-state effects implemented in the AMPT model. The correlation function is an intuitive observable for understanding the influence of different underlying physics.

Note that the correlation function obtained via this procedure is affected by a small bias from short-range component, denoted as  $\delta_{\mathrm{SRC}}(\eta_1,\eta_2)$, via the normalization procedure of Eq.~\ref{eq:13}. The $\delta_{\mathrm{SRC}}(\eta_1,\eta_2)$ distribution typically is relatively flat along $\eta_1+\eta_2$ with a rather narrow width in the $\eta_1-\eta_2$ direction. In this case, one can easily see that the contribution of $\delta_{\mathrm{SRC}}(\eta_1,\eta_2)$ to $C_p$ is not uniform in $\eta$: if the first particle is near mid-rapidity $\eta_1\sim0$ then all pairs in $\delta_{\mathrm{SRC}}(\eta_1,\eta_2)$ contributes to $C_p(\eta_1)$, whereas if the first particle is near the edge of the acceptance $\eta_1\sim\pm Y$ then only half of the pairs in $\delta_{\mathrm{SRC}}(\eta_1,\eta_2)$ contributes to $C_p(\eta_1)$. However the short-range component contribution can be estimated, e.g. via an experimental procedure discussed in Ref.~\cite{Jia:2016jlg}, then such acceptance bias can be removed by redefining the projection function and $C_N$ function as:
\begin{eqnarray}
\label{eq:c2b}
&&C_{p}^{\rm{sub}}(\eta_1) = \frac{\int \left[C(\eta_1,\eta_2)-\delta_{\mathrm{SRC}}(\eta_1,\eta_2)\right] d\eta_2}{2Y},\;C_{p}^{\rm{sub}}(\eta_2) = \frac{\int \left[C(\eta_1,\eta_2)-\delta_{\mathrm{SRC}}(\eta_1,\eta_2)\right] d\eta_1}{2Y}\;,\\\label{eq:c2c}
&&C_{\rm N}'(\eta_1,\eta_2) = \frac{C(\eta_1,\eta_2)}{C_{p}^{\rm{sub}}(\eta_1)C_{p}^{\rm{sub}}(\eta_2)}.
\end{eqnarray}
Therefore $C_{\rm N}'$ is only corrected for the residual centrality dependence and is free of bias from short-range correlations. One can use $C_{\rm N}'$ instead of $C_{\rm N}$ to extract $a_n$-spectra. The main effect of the bias is reduce the value of $C_{\rm N}$ relative to $C_{\rm N}'$ at the four corner's of the $\eta_1,\eta_2$ phase space. We shall leave this topic for a future study.

\section{discussion and summary}
\label{sec:dis}
We have introduced two complimentary methods for detailed study of the event-by-event fluctuations of particle production in the longitudinal direction. The single-particle method gives the coefficients in each event, which can be directly relate to the fluctuation of the initial geometry in model calculation. On the other hand, two-particle correlation method suppresses the statistical noise on the ensemble basis and hence does not require the construction of random events. The correlation method is particularly suitable for small collision system, such as $p+p$ or $p$+Pb collisions, where the EbyE statistical fluctuation is very large. Furthermore, the influence of the detector effects is straightforward to remove in the correlation method, and hence it should be considered as the primary method in the experimental data analysis.

\begin{figure}[!t]
\begin{center}
\includegraphics[width=0.85\columnwidth]{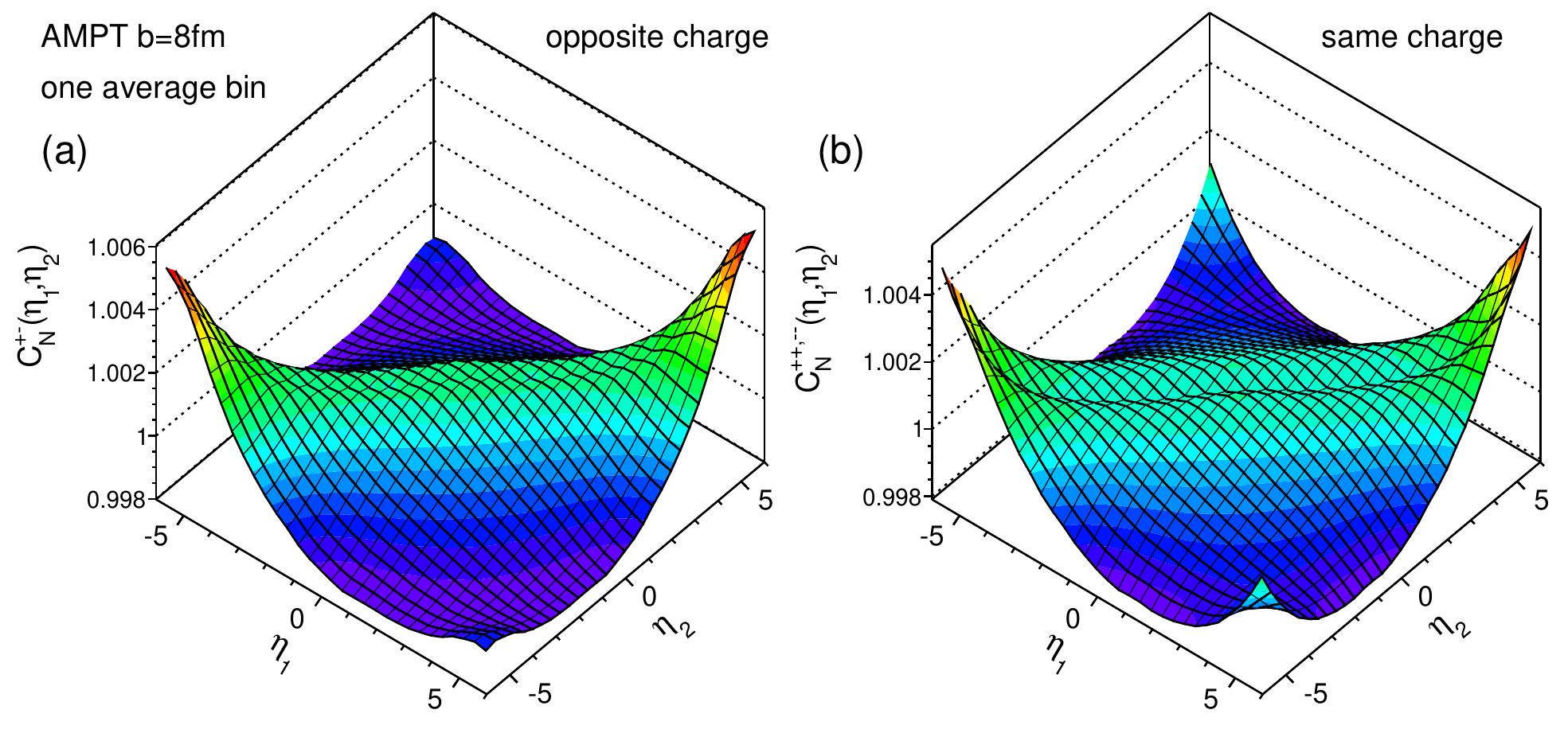}
\end{center}
\caption{\label{fig:10}  The correlation functions for same-charge pairs (left panel) and opposite-charge pairs (right panel) for AMPT events generated with $b=8$~fm.}
\end{figure}

The correlation method discussed in this paper can be generalized into correlation between multiplicity of particles of any two different types. For example one can measure the correlation between multiplicities for positive and negative particles:
\begin{eqnarray}
\label{eq:16}
C^{+-}(\eta_1,\eta_2) &=& \frac{\left\langle N^+(\eta_1) N^-(\eta_2)\right\rangle}{\left\langle N^+(\eta_1)\right\rangle\left\langle N^-(\eta_2)\right\rangle},
\end{eqnarray}
which allow the extraction of $\lr{a_n^+a_n^-}$. Assuming equal multiplicity for positive and negative particles, the coefficients for positive particle $a_n^+$ and negative particles $a_n^-$ are related to those for inclusive particles via:
\begin{eqnarray}
\label{eq:17}
\lr{a_n^2} = \frac{1}{4}\lrp{\lr{a_n^+a_n^+}+ \lr{a_n^-a_n^-}+2\lr{a_n^+a_n^-}}
\end{eqnarray}
Due to local charge conservation effects, the correlation between positive and negative particles is expected to be stronger than inclusive correlation. Indeed the AMPT or HIJING simulation studies suggest that $\lr{a_n^+a_n^-}>\lr{a_n^2}>\lr{a_n^+a_n^+}=\lr{a_n^-a_n^-}$. The results shown in Fig.~\ref{fig:10} implies that the dip around $\eta_1\sim \eta_2$ seen in the inclusive correlations for AMPT model (e.g. Fig.~\ref{fig:7a}) arises mainly from same-charge pairs, although the opposite-charge pair correlation also shows a shallow dip. Such dip is absent in HIJING events independent of the charge combination. These structures reflect the important role of the final-state interaction and hardronization mechanism (via simple coalescence in AMPT) on the charge-dependent correlations. Note that the charge-dependent correlation function is related to the well known balance function $B(\Deta)$~\cite{Bass:2000az}:
\begin{eqnarray}
\label{eq:18}
 2B(\Deta) = 2C^{+-}(\Deta)- C^{++}(\Deta)- C^{--}(\Deta),
\end{eqnarray}
The stronger correlation strength for opposite-charge pairs than the same-charge pairs as shown in Fig.~\ref{fig:10}, implies that the balance function should peak around $\Deta=\eta_1-\eta_2=0$ and fall slowly to large $\Deta$ (i.e. not sensitive to the dips), consistent with earlier observations~\cite{Adams:2003kg,Abelev:2013csa}. 

Similarly, one could also divide particles into high $\pT$ and low $\pT$ with equal multiplicity. In this case, the coefficients can be written as 
\begin{eqnarray}
\label{eq:19}
\lr{a_n^2} \approx \frac{1}{4}\lrp{\lr{a_n\su{H}a_n\su{H}}+ \lr{a_n\su{L}a_n\su{L}}+2\lr{a_n\su{H}a_n\su{L}}}
\end{eqnarray}
where $a_n\su{H}$ and $a_n\su{L}$ are coefficients for high $\pT$ and low $\pT$ particle multiplicity, respectively (for example $>1$ GeV/$c$ and $<1$ GeV/$c$). We observe that $\lr{a_n\su{H}a_n\su{H}}>\lr{a_n\su{H}a_n\su{L}}>\lr{a_n\su{L}a_n\su{L}}$ (not shown), presumably due to short-range correlations related to jet fragmentation, which are stronger for higher $\pT$ particles. It would be interesting to study the factorization behavior of the multiplicity correlation by calculating a factorization ratio, similar to what is often used in azimuthal flow correlation analysis~\cite{Gardim:2012im}:
\begin{eqnarray}
\label{eq:20}
r_n = \frac{a_n\su{H}a_n\su{L}}{\sqrt{\lr{a_n\su{H}a_n\su{H}}}\sqrt{\lr{a_n\su{L}a_n\su{L}}}}
\end{eqnarray}
The breaking of the factorization can be used to understand the $\pT$ dependence of the long-range and short-range correlations.

The $a_n$ coefficients can be significantly affected by the short-rangle correlations. One way to suppress such short-range correlation is by requiring the pairs to be separated in azimuthal angle $\phi$~\cite{Vechernin:2013vpa,Adam:2015mya}~\footnote{In principle, the full information of the transverse and longitudinal multiplicity and flow fluctuations is contained in the 3-D correlation function $C(\eta_1,\eta_2,\Dphi)$.}.  However the challenge is to understand role of the harmonic flow $v_n$ and their EbyE fluctuations, since harmonic flow introduces non-trivial multiplicity correlations between particles in different $\phi$ regions.

In order to study dependence of observables on the size of the collision system, many measurements classify collisions according to event activity or centrality in certain $\eta$ range. The key challenge in centrality definition is to understand dynamical multiplicity correlations between the $\eta$ range used for centrality determination and $\eta$ range used for the observable. This is an open issue particularly important in small collision system such as $p+p$ and $p$+Pb collisions, where the bias associated with centrality selection often dominates over the experimental uncertainties~\cite{Jia:2009mq,Adare:2013nff,ATLAS-CONF-2013-096,Adam:2014qja,Perepelitsa:2014yta}. Our method can be used to measure and quantify such multiplicity correlations, which can then be used to understand the influence of centrality biases in other measurements. Since $p$+Pb is an asymmetric collision system, the correlations between odd and even terms may not vanish, which can be studied by measuring $\lr{a_na_{n+1}}$. 

In summary, a method has been proposed to study the longitudinal multiplicity correlations in high-energy nuclear collisions. In this method, events are classified into narrow event activity bins, and EbyE fluctuations are then extracted relative to the average multiplicity distribution in each event activity bin. This procedure allows the separation of the centrality dependence of the multiplicity distribution from the dynamical shape fluctuations. The multiplicity correlations are extracted using the single-particle distribution or two-particle correlation function. The extracted signals are decomposed into a set of orthogonal longitudinal harmonics in terms of Legendre polynomials, which characterize various components of the multiplicity fluctuation of difference wavelength in $\eta$. The first several coefficients $a_n$ are obtained and found to decrease slowly with $n$ in HIJING model but very rapidly with $n$ in AMPT model, which could be due to viscous damping effects of the longitudinal harmonics by the final-state rescattering effects. The $a_1$ signal is found to strongly correlated with the asymmetry in the number of forward-going and backward-going participating nucleons; while a nonzero $a_2$ signal could be related to the fluctuations of the nuclear stopping or shift of the effective center-of-mass of the collisions. This geometrical origin of the $a_1$ can be experimentally verified by observing an anti-correlation between $a_1$ and the asymmetry of the spectator nucleons detected by the zero-degree calorimeters. Two-particle pseudorapidity correlations also reveal interesting charge-dependent short-range structures in AMPT model but are absent in HIJING model, suggesting that these structures are sensitive to the underlying hadronization mechanism. Hence measurement of the multiplicity fluctuation in terms of longitudinal harmonics provide an promising avenue for understanding the particle production mechanism in the early stage of the heavy-ion collisions and for probing the final-state rescattering effects. The proposed two-particle correlation method is particularly suitable for high-energy proton-lead and proton-proton collisions where the longitudinal multiplicity fluctuations are very large and are responsible for the biases in the centrality definition. Since our method correlates event activities between separate rapidity ranges, it provides a useful way to unfold and quantify the centrality correlations between different rapidity ranges.

We appreciate fruitful discussions with R.~Lacey. This research is supported by NSF under grant number PHY-1305037 and by DOE through BNL under contract number DE-SC0012704.

\bibliography{FBmulv2}{}
\bibliographystyle{apsrev4-1}

\end{document}